\documentclass[aps,reprint,superscriptaddress,nofootinbib,showpacs,amsmath,amssymb]{revtex4-1}
\usepackage{latexsym,graphicx,slashed,bm,dcolumn}

\newcommand{\mat}[4]{\left(\begin{array}{cc}{#1}&{#2}\\{#3}&{#4}
\end{array}\right)}

\newcommand{\ov}{\overline}

\newcommand*\xbar[1]{%
\kern0.5ex%
 \hbox{%
  \kern0.05ex%
     \vbox{%
     \hrule height 0.5pt % The actual bar
     \kern0.25ex%         % Distance between bar and symbol
     \hbox{%
       \kern-0.1em%      % Shortening on the left side
       \ensuremath{#1}%
       \kern-0.1em%      % Shortening on the right side
     }%
   }%
 }%
}
\begin{document}

%\qquad\qquad\qquad FTPI-MINN-15/29, NSF-KITP-15-073

\title{Neutron--Antineutron Oscillations: Discrete Symmetries and Quark Operators}

\author{Zurab Berezhiani} 
\affiliation{Dipartimento di Fisica e Chimica, Universit\`a dell'Aquila, Via Vetoio, 67100 Coppito, L'Aquila, Italy} 
\affiliation{INFN, Laboratori Nazionali del Gran Sasso, 67010 Assergi,  L'Aquila, Italy}

\author{Arkady Vainshtein}
\affiliation{School of Physics and Astronomy and William I. Fine Theoretical Physics Institute, University of Minnesota,
Minneapolis, MN 55455, USA}
\affiliation{Kavli Institute for Theoretical Physics, University of California, Santa Barbara, CA 93106, USA}

%\date	{}						% Activate to display a given date or no date

\begin{abstract}
We analyze status of {\bf C}, {\bf P} and {\bf T} discrete symmetries in application to  neutron-antineutron transitions 
breaking conservation of baryon charge ${\cal B}$ by two units. At the level of free particles all these symmetries are preserved.
This includes  {\bf P} reflection in spite of the opposite internal parities usually ascribed to neutron and antineutron. 
Explanation, which goes back to the 1937 papers by E.~Majorana and by G.~Racah, is based on a definition of parity satisfying ${\bf P}^{2}=-1$,
instead of ${\bf P}^{2}=1$,
and ascribing  $ {\bf P}=i$ to both, neutron and antineutron. 
We apply this to {\bf C}, {\bf P} and {\bf T} classification of six-quark operators with $|\Delta {\cal B} |=2$. It allows to
specify operators contributing to neutron-antineutron oscillations. Remaining  operators contribute to other
$|\Delta {\cal B} |=2$ processes and, in particular, to nuclei instability.
We also show that presence of external magnetic field does not induce any new operator
mixing the neutron and antineutron provided that rotational invariance is not broken.
 \end{abstract}

\maketitle

\noindent
{\bf 1.}  A phenomenon of neutron-antineutron oscillation was suggested by Kuzmin~\cite{Kuzmin:1970nx}
in 1970, and the first theoretical model -- by Mohapatra and  Marshak in 1980~\cite{Mohapatra:1980qe}. 
It is now under active discussion (for a review, see~\cite{Phillips:2014fgb}). 
A discovery of this oscillations would be a clear evidence of baryon charge nonconservation, 
$|\Delta {\cal B} |=2$.
In this note we discuss the issue of {\bf C}, {\bf P} and {\bf T} symmetries 
in the $|\Delta {\cal B} |=2$ transitions, applying this to analysis of six-quark operators.
We also analyze effects of external magnetic field 
and show that it does not add any new $|\Delta {\cal B} |=2$ operator if the rotational invariance is not broken.

Essentially the same issues were addressed in our previous note~\cite{BV}. There we emphasize the point
that parity {\bf P}, defined in such a way that ${\rm \bf P}^{2}\!=\!1$,  is broken, as well as  {\bf CP},  in 
the neutron-antineutron transition. This is an immediate consequence 
of the opposite parities of neutron and antineutron when  ${\rm \bf P}^{2}\!=\!1$. Indeed, we deal then with mixing of the states with different parities. 
Although we also noted that in the absence of interaction it does not automatically imply 
an existence of CP breaking physics we did not present a detailed analysis of the problem. 
We have corrected this at the INT workshop in September 2015, 
defining ${\rm\bf P}_{\!z}$ such that ${\rm\bf P}_{\!z}^{\,2}=-1$. 

Following our note~\cite{BV} the issue of parity definition in the  $|\Delta {\cal B} |=2$ transitions 
was addressed in a number of related publications \cite{Fujikawa,McKeen,Gardner}. 
Unfortunately, together with correct statements some of these analyses are clearly erroneous. 
For instance, McKeen and Nelson  in their interesting paper \cite{McKeen}  about CP violation
due to baryon oscillations wrongly insisted that one can keep ${\rm \bf P}^{2}\!=\!1$ for the parity definition.
It shows that the subject deserves a further discussion. 
Actually, the issue of parity definition for fermions was resolved 
long ago.
Below we present more details of parity definition story which has been started in 1937 
by Ettore Majorana in his famous paper  \cite{Majorana} where he introduced a notion of Majorana fermions. 
In the same journal issue the parity definition was discussed in more details 
 by Giulio Racah \cite{Racah}. 
%\\[1mm]

\noindent
{\bf 2.}   Let us start with the Dirac Lagrangian
\begin{equation}
{\cal L}_{D}=i\xbar n \gamma^{\mu} \partial_{\mu}n - m\xbar n \,n
\label{lagr}
\end{equation}
with the four-component spinor $n_{\alpha}\,,~(\alpha=1,...,4)$ and the mass parameter $m$ which is real and positive.  
The Lagrangian
 gives the Lorentz-invariant description
of free neutron and antineutron states and preserves
the baryon charge, ${\cal B}=1$ for $n$ and ${\cal B}=-1$
for $\xbar n$\,. Its conservation is associated with the continuous ${\rm U}(1)_{\cal B}$ symmetry
\begin{equation}
n \to {\rm e}^{i\alpha }n , \quad \xbar n \to {\rm e}^{-i\alpha }\xbar n 
\label{U1B}
\end{equation}
of Lagrangian (\ref{lagr}). Correspondingly, at each spatial momentum there are four degenerate states,
the spin doublet of the neutron states with the baryon charge ${\cal B}=1$, and the spin doublet
of the antineutron states with ${\cal B}=-1$, 
i.e.,  two spin doublets which differ by the baryon charge ${\cal B}$. 

Note that another bilinear mass term,
\begin{equation}\label{lagrprime}
 - i  \widetilde m \, \xbar n \gamma_{5} n\, ,
\end{equation}
consistent with the baryon charge conservation, 
can be rotated away by chiral ${\rm U}(1)$ transformation $n\to  {\rm e}^{i\eta \gamma_{5}}n$\,.

How the baryon number non-conservation shows up at the level of free one-particle 
states? In Lagrangian description it could be only modification of the bilinear mass terms.
Generically, there are four such Lorentz invariant bilinear terms:
\begin{equation}
n^{T}\!Cn\,,\quad n^{T}\!C\gamma_{5}n\,,\quad \xbar n\, C {\xbar n}^{T},\quad \xbar n \,C \gamma_{5}{\bar n}^{T}.
\label{mod}
\end{equation}
Here $C=i\gamma^{2}\gamma^{0}$ is the charge conjugation matrix 
in the Dirac (standard) representation of gamma matrices. It has the same form in the Weyl (chiral) representation.
In the Majorana representation $C=-\gamma^{0}$.

Using the chiral basis we show in the part 4  that all these modifications (\ref{mod}) are reduced by field redefinitions to 
just one possibility for the baryon charge breaking  by two units,
\begin{equation}
\Delta {\cal L}_{\not {\cal B}}=-\frac{1}{2} \,\epsilon \,\big [n^{T}\!Cn +\xbar n\, C \!{\xbar n}^{T}\,\big ]\,,
\label{deltaB}
\end{equation}
where $\epsilon$ is a real positive parameter. The possibility of such redefinitions is based on U(2) symmetry 
of the kinetic term $i\xbar n \,\gamma^{\mu} \partial_{\mu}n$.  Four-parametric U(2) transformations allow to exclude the term (\ref{lagrprime}) and to reduce four terms (\ref{mod}) to just one structure (\ref{deltaB}).
\\[1mm]

\noindent
{\bf 3.}  What is the status of discrete 
{\boldmath${\rm C,\,P}$} and {\bf T} symmetries under the baryon charge breaking modification (\ref{deltaB})?
Let us first consider the charge conjugation {\bf C}, which can be viewed as a plain exchange symmetry between $n$ and
$n^{c}$ fields,
\begin{equation}
{\rm \bf C}:\quad n \longleftrightarrow n^{c}=C{\xbar n}^{T}\,.
\label{conj}
\end{equation}
This is a sort of discrete $Z_{2}$ symmetry, ${\rm \bf C}^{2}=1$. The most simple it looks in the Majorana representation where
\begin{equation}
 n^{c}=n^{*}\,.
\label{conjM}
\end{equation}

It is straithforward to verify that both Lagrangians above, (\ref{lagr}) and (\ref{deltaB}),
are {\bf C} invariant. Indeed, they could be rewritten in the form
\begin{equation}
\begin{split}
&{\cal L}_{D}=\frac{i}{2}\big[\xbar n \gamma^{\mu} \partial_{\mu}n+ \xbar {n^{c}} \,\gamma^{\mu} \partial_{\mu}n^{c}\big]
 - \frac{m}{2}\big[\xbar n\, n+\xbar {n^{c}}\,n^{c}\big],\\
&\Delta {\cal L}_{\not{\cal B}}~=-\frac{\epsilon}{2} \, \big[\!\xbar{n^{c}}\, n +\xbar n \, n^{c}\big],
\end{split}
\end{equation}
which makes their {\bf C} invariance explicit.

The Lagrangians are diagonalized in terms of Majorana fields $n_{1,2}$\,,
\begin{equation}
n_{1,2}=\frac{n\pm n^{c}}{\sqrt{2}}
\,,
\label{Maj}
\end{equation}
which are even and odd under the charge conjugation {\bf C}, $n^{c}_{1,2}= \pm \,n_{1,2}$.
Namely,
\begin{equation}
\begin{split}
&{\cal L}_{D}=\frac{1}{2}\sum_{k=1,2}\big[\xbar n_{k} \gamma^{\mu} \partial_{\mu}n_{k}
-m\xbar n_{k} n_{k}\big],\\
&\Delta {\cal L}_{\not{\cal B}}~=-\frac{1}{2} \,\epsilon \big[\xbar{n}_{1}\, n_{1} -\xbar n_{2} \, n_{2}\big].
\end{split}
\end{equation}
It demonstrates that the baryon charge breaking leads to splitting into two Majorana spin doublets.
The C-even $n_{1}$ field gets the mass $M_{1}=m+\epsilon$ while the mass of the C-odd $n_{2}$ is $M_{2}=m-\epsilon$.

Turn now to the parity transformation {\bf P}. It involves (besides reflection of the space coordinates)
the substitution 
\begin{equation}
{\rm \bf P}:\quad n \to \gamma^{0}n\,,\qquad n^{c} \to -\gamma^{0}n^{c}\,,
\label{parity}
\end{equation}
 where $\gamma^{0}C\gamma^{0}=-C$ is used. The opposite signs in transformations for $n$ and $n^{c}$
 reflect the well-known theorem\,\cite{Berest} on the opposite parities of fermion and antifermion. The definition (\ref{parity}) satisfies ${\rm \bf P}^{2}=1$,
 so the eigenvalues of {\boldmath${\rm P}$} are $\pm 1$ and opposite for fermion and antifermion states.
 
 Different parities of neutron and antineutron imply that their mixing breaks {\bf P} parity, and, indeed,
 the substitution (\ref{parity}) changes $\Delta {\cal L}_{{\not\cal B}}~$ to $(-\Delta {\cal L}_{{\not\cal B}})$\,.
Together with {\bf C} invariance it implies then that $\Delta {\cal L}_{\not{\cal B}}$\, is also {\bf CP} odd.
However, this {\bf CP} oddness does not translate immediately into observable {\bf CP} breaking effects. 
To get them one needs an interference of amplitudes and this is provided only when interaction is present.

It shows a subtlety in the definition of parity transformation {\bf P}, see textbook discussions, e.g., in Refs.\,\cite{BLP,PeskSch}.
Let us remind it. 

 When baryon charge is conserved there is no transition between sectors with different ${\cal B}$,
 and one can combine {\bf P} with a baryonic ${\rm U}(1)_{\cal B}$ phase rotation (\ref{U1B}) and define {\boldmath${\rm P}_{\alpha}$},
\begin{equation}
{\rm \bf P}_{\alpha}={\rm \bf P}\,{\rm e}^{i{\cal B}\alpha}: \quad n \to {\rm e}^{i\alpha}\gamma^{0}n\,,
\quad n^{c} \to -{\rm e}^{-i\alpha}\gamma^{0}n^{c}\,.
\label{palpha}
\end{equation}
Of course, then ${\rm \bf P}_{\alpha}^{2}={\rm e}^{2i{\cal B}\alpha}\neq 1$ but the phase is unobservable when 
${\cal B}$ is conserved.

When baryon charge is not conserved the only remnant of baryonic ${\rm U}(1)_{\cal B}$ rotations
is $Z_{2}$ symmetry associated with changing sign of the fermion
field, $n\to -n$. This symmetry is protected: unphysical $2\pi$ space rotation
changes the sign of the fermion field.
It means that besides the original ${\rm \bf P}^{\,2}=1$ we can consider  
a different parity definition ${\rm \bf P}_{z}$\,, such that  ${\rm \bf P}_{z}^{2}=-1$.

Thus, choosing $\alpha= \pi/2$ in Eq.\,(\ref{palpha}), we come to a new parity
${\rm \bf P}_{z}$,
\begin{equation}
{\rm \bf P}_{z}={\rm \bf P}\,{\rm e}^{i{\cal B}\pi/2}: \quad n \to  i\gamma^{0}n\,,
\quad n^{c} \to i\gamma^{0}n^{c}
\label{pz}
\end{equation}
with ${\rm \bf P}_{z}^{2}=-1$.
Now ${\rm \bf P}_{z}$ parities of $n$ and $n^{c}$ states are the same and equal to $i$,
so their mixing does not break ${\rm \bf P}_{z}$ parity. It means that 
all discrete symmetries, {\bf C}, ${\rm \bf P}_{z}$ and {\bf T} are preserved by the baryon breaking term 
$\Delta {\cal L}_{\not{\cal B}}$\,.

Couple of related comments. First, one can choose $\alpha\!=\! -\pi/2$ and have parities of  fermion and antifermion 
both equal to $(-i)$ instead of $i$.
The absolute sign has no physical meaning -- it could be changed by a $2\pi$ space rotation -- but relative parity between two different fermions does make sense. 
Second, it is amusing that the same ${\rm \bf P}_{z}$ parity for $n$ and $n^{c}$ equal to $i$ is still consistent with the notion of 
opposite parities of fermion and antifermion, having in mind that that for the complex value of parity 
we should compare ${\rm \bf P}_{z}(n)$ with $[{\rm \bf P}_{z}(n^{c})]^{*}$. Also for a fermion-antifermion pair the product  
${\rm \bf P}_{z}(n) {\rm \bf P}_{z}(n^{c})=-1$. One more comment is to notice that  ${\rm \bf P}_{z}$ commutes with {\bf C}, i.e.,
{\bf C}${\rm \bf P}_{z}$=${\rm \bf P}_{z}${\bf C}, in contrast with
{\bf P} which instead anticommutes with {\bf C}, i.e., {\bf C}{\bf P}=$-${\bf P}{\bf C}\,. For Majorana fermion both charge and parity conjugations are diagonal in the Hilbert space: their actions (in the rest frame) do not lead to a different physical state. It means that only the commuting case, i.e.,  ${\rm \bf P}_{z}$ not {\bf P}, is allowed.

Thus, we demonstrated that neutron-antineutron mixing by $\Delta {\cal B}=\pm 2$ Majorana term in the mass matrix 
leads to a specific definition of the conserved parity ${\rm \bf P}_{z}$, making it complex and satisfying 
${\rm \bf P}_{z}^{2}=-1$ instead of (+1). It is this definition which should be used in analyzing ${\bf CP_{z}}$ violating
interactions. 

Having in mind that invariance under the charge conjugation was already checked, 
preservation of {\bf T} invariance follows from {\bf CPT} theorem provided by Lorentz invariance and locality. A specific ${\rm \bf P}_{z}$ definition of parity transformation defines a specific {\bf T} transformation.

A few words about the history of the parity definition. As we mentioned earlier Ettore Majorana and Giulio Racah were 
the first to realize a necessity of ${\rm \bf P}_{z}^{2}=-1$ in application to Majorana fermions \cite{Majorana, Racah}.
The case of neutron-antineutron mixing is essentially the same because it leads to splitting into two Majorana fermions of different mass and opposite {\bf C}-parities. 
Years later this definition of parity was applied to Majorana neutrino in Refs.\,\cite{Wolfenstein}.

Let us now comment on the recent applications \cite{Fujikawa,McKeen,Gardner} to neutron-antineutron oscillations. 
Fujikawa and Tureanu in \cite{Fujikawa} came to incorrect conclusion about necessity of  {\bf P}-parity breaking in $\Delta {\cal B}\!=\!\pm 2$ processes. Similar to our initial claim in \cite{BV}, it is due ${\rm \bf P}^{2}=1$ for the parity definition what leads to the opposite parities $\pm 1$ for neutron and antineutron.
 McKeen and Nelson in \cite{McKeen} also missed this point, and, as we mentioned at the beginning, incorrectly insisted that one can stay with ${\rm \bf P}^{2}\!\!=\!1$. Technically, the origin of the mistake is that their $\Delta {\cal B}\!=\!\pm 2$ Lagrangian, given by Eq.\,(A6) in \cite{McKeen}, becomes a total derivative when {\bf C} and {\bf P} are conserved with ${\rm \bf P}^{2}=1$. 
 Then, all its matrix elements vanish - no oscillations. 
 Gardner and Yan in \cite{Gardner}  followed Majorana neutrino case \cite{Wolfenstein} and correctly defined  the parity inversion with ${\bf P}_{z}^{2}\!=\!-1$ and the same ${\rm \bf P}_{z}\!=\!i$ for botn, neutron and antineutron.
\\[1mm]

\noindent
{\bf 4.} To show that the above consideration covers a generic case it is convenient
to introduce two left-handed Weyl spinors, forming a flavor doublet\footnote{See, e.g., the book~\cite{Ramond:1999vh}
where the Weyl spinor formalism is gracefully applied to description of massive neutrinos.} 
\begin{equation}
\psi^{i\,\alpha}= \left(\begin{array}{c} \psi^{1\alpha}  \\[1mm]  \psi^{2\alpha}  \end{array}\right) ,
\qquad i=1,2, \quad \alpha=1,2\,,
\label{doubl}
\end{equation}
together with their complex conjugates,  
\begin{equation}
\xbar{\psi}^{\,\dot\alpha}_i\equiv \Big( (\psi^{1\dot \alpha})^{*},(\psi^{2\dot \alpha})^{*}\Big )\,,\qquad i=1,2,\quad \dot\alpha=1,2\,,
\label{doubl-r}
\end{equation}
representing the right-handed spinors.
One can raise and lower spinor $\alpha, \dot\alpha$ and flavor $i$ indices using $\epsilon_{\alpha\beta}\,,\,\epsilon_{\dot\alpha\dot\beta}$
and $\epsilon_{ik}$ with $\epsilon^{12}=1$.

In terms of Dirac spinor $n$  two left-handed Weyl spinors (\ref{doubl}) are associated with 
$n_{L}$ and $n^c_L= i\gamma^2 (n_{R})^{*}$.
In particular, in the chiral (Weyl) basis of gamma-matrices we have: 
\begin{equation}
n=\left(\!\!\begin{array}{c}\psi^{1}\\[1mm] -i\sigma^{2}(\psi^{2})^{\ast} \end{array}\!\!\!\right)=
\left(\begin{array}{c}\psi^{1\alpha}\\[1mm] \bar\psi_{2\dot \alpha} \end{array}\right), 
\quad n^{c}=\left(\begin{array}{c}\psi^{2\alpha}\\[1mm] \bar\psi_{1\dot \alpha} \end{array}\right).
\label{dirac}
\end{equation}

The generic Lorentz invariant Lagrangian quadratic in fermionic fields $\psi^{i\alpha}$ 
and $\xbar{\psi}_{i\dot\alpha}$ is
\begin{equation}
\begin{split}
& {\cal L} = 
 \frac{i}{2} \, \Big( \psi^k_\alpha \, \partial^{\alpha\dot\alpha} \xbar{\psi}_{k\dot\alpha} \!+ \!
\xbar{\psi}_k^{\,\dot\alpha} \, \partial_{\dot\alpha \alpha} \psi^{k\alpha} \Big) 
 \\[1mm]
&
\quad  -\frac 12\, \Big(m_{ik}\, \psi^i_\alpha \, \psi^{k\alpha} 
 +\xbar{m}^{\,ki} \, \xbar{\psi}_k^{\,\dot\alpha} \, \xbar{\psi}_{i\dot\alpha}\Big)\,,
 \end{split}
\label{weyl}
\end{equation}
where 
$\partial^{\alpha\dot\alpha}\!=\! (\sigma^{\mu})^{\alpha\dot\alpha}\partial_{\mu}$, 
$\sigma^{\mu}\!=\!\{1,\vec \sigma\}$,  
and $\partial_{\dot\alpha \alpha} \!=\! (\bar\sigma^{\,\mu})_{\dot\alpha\alpha}\partial_{\mu}$\,, 
$\bar\sigma^{\mu}\!=\!\{1,-\vec \sigma\}$,  
$m_{ik}$ is the symmetric mass matrix, $m_{ik}\!=\!m_{ki}$ and 
$\xbar{m}^{\,ik}=(m_{ik})^{*}$ is its conjugate. 

In the above equation we are implying a standard diagonal form for kinetic terms.  
These terms in (\ref{weyl}) are U(2) symmetric: besides flavor SU(2) rotations it includes also
U(1) associated with the overall phase rotation of the flavor doublet (\ref{doubl}) which in terms 
of Dirac spinors (\ref{dirac}) is just a chiral transformation. 
The U(2) symmetry of kinetic terms it clearly generic: starting with 
$i\xbar{\psi}^{\,\dot\alpha}_i C^{i}_{k}\,\partial_{\alpha\dot\alpha} \,\psi^{k\alpha}$ where $C^{i}_{k}$ is an arbitrary  Hermitian flavor matrix,
one can always diagonalize and normalize these terms.

As for the mass terms they generically break both, U(1) and SU(2) flavor symmetries, so no 
continuous symmetry remains. 
To see how the U(1)$_{\cal B}$ symmetry (\ref{U1B}) associated with the baryon charge
could survive note that one can interpret U(2) transformations as acting on the external 
mass matrix $m_{ik}$. This matrix is charged under U(1), the overall phase rotation, so
this U(1) symmetry is always broken by nonvanishing mass. In respect to SU(2) transformations the symmetric
tensor $m_{ik}$ is the adjoint representation, i.e., can be viewed as an isovector $\mu^{a}$, $a=1,2,3$,
\begin{equation}
m^i_k=\varepsilon^{ij} m_{jk} =  \mu^a (\tau^a)^i_k\,,\quad a=1,2,3\,, 
\label{defmu}
\end{equation}
Because $\mu^{a}$ is complex, we are actually dealing with two real isovectors,
 ${\rm Re}\,\mu^{a}$ and  ${\rm Im}\,\mu^{a}$. 
The SU(2) transformations are equivalent to simultaneous rotation of both
vectors, while U(1) changes phases of all $\mu^{a}$ simultaneously, which is equivalent  
to SO(2)  rotation inside each couple $\{{\rm Re}\,\mu^{a}$, ${\rm Im}\,\mu^{a}\}$. 
 Only in case when these 
 vectors are parallel we have an invariance of the mass matrix which is just
 a rotation around this common direction. This symmetry is the one
 identified with the baryonic ${\rm U}(1)_{\cal B}$ in Eq.\,(\ref{U1B}).
When it happens all ${\rm Im}\,\mu^{a}$ can be absorbed in ${\rm Re}\,\mu^{a}$ by 
 U(1) transformation. 
 
Let us show now that in the absence of the common direction we get 
two spin 1/2 Majorana fermions with different masses.
From equations of motion
\begin{equation}
\begin{split}
&i\,\partial_{\dot\alpha\alpha} \psi^{i\alpha} - 
\xbar{m}^{\,ik}\xbar{\psi}_{k\dot\alpha}\!=0\,,\\[1mm]
& i\,\partial^{\alpha\dot\alpha}\xbar{\psi}_{i
  \dot\alpha}-m_{ik}\,\psi^{k \alpha}=0~,
\end{split}
\end{equation}
we come to the eigenvalue problem for
$M^2=p_\mu p^\mu$,
\begin{equation}
M^2\psi^{k\alpha}-\xbar{m}^{\,ki}m_{il}\, \psi^{l\alpha}=0\,.
\end{equation}

Using definition (\ref{defmu}) of $\mu^{a}$ the squared mass matrix 
can be presented as a combination of isoscalar and isovector pieces:  
\begin{equation}
\xbar{m}^{kl}m_{ln}=\mu^a\xbar{\mu}^{\,a}\,\delta^k_n  +i\epsilon^{abc}\mu^a\xbar{\mu}^{\,b} (\tau^c)^k_n\,.
\label{mu2}
\end{equation}
Correspondingly, there are two invariants defining $M^2$. The isoscalar part gives the sum of eigenvalues,
\begin{equation}
\frac{M_1^2+M_2^2}{2}=\mu^a\xbar{\mu}^{\,a}= ({\rm Re} \,\mu^a)^2+({\rm \,Im}\,\mu^a)^2\, 
\end{equation}
while the length of the isovector part defines the splitting of the eigenvalues,
\begin{equation}
\frac{M_1^2-M_2^2}{2}=2\sqrt{\Big[\epsilon^{abc} \,{\rm Re} \,\mu^a {\rm \,Im}\,\mu^b\Big]^2}\,.
\end{equation}
This shows how the splitting is associated with the breaking of the baryon charge.

To follow the discrete symmetries we can orient the  mass matrix $m_{ik}$ in a convenient way.
In terms of $\mu^{a}$ the mass matrix $\widehat m$ has the form
\begin{equation} 
\widehat m=m_{ik} = \mat{-\mu^{1}- i\mu^{2} }{\mu^{3} }{\mu^{3}}{\mu^{1} - i\mu^{2}}  . 
\label{mass-matrix}
\end{equation} 
Without lost of generality one can render the vectors ${\rm
  Re}\,\mu^{a}$ and  ${\rm Im}\,\mu^{a}$ orthogonal using the overall U(1)
phase transformation. Then by remaining SU(2) rotations we can put
both of them  onto the 23 plane,
i.e., put $\mu^{1}=0$. and choose the direction of ${\rm Im}\,\mu^a$
as the 2-nd axis.
So, only two non-vanishing parameters, ${\rm Re}\,\mu^{3}$ and ${\rm Im}\,\mu^{2}$, remain
and the mass matrix takes the form
\begin{equation} 
\widehat m_{0} = \mat{{\rm Im}\,\mu^{2} }{{\rm Re}\,\mu^{3} }{{\rm Re}\,\mu^{3}}{{\rm Im}\,\mu^{2}} 
=\mat{\epsilon}{m }{m}{\epsilon},
\label{mass-matrix1}
\end{equation} 
where correspondence, $m\!=\!{\rm Re}\,\mu^{3}$,~$ \epsilon\!=\!{\rm Im}\,\mu^{2}$. with parameters introduced earlier in four-component spinor notations is also shown. 
Then $M_{1,2}^{2}=(m\pm \epsilon)^{2}$ as in the previous Section.

In other words, an arbitrary mass matrix $\widehat{m}$, as in Eq.\,(\ref{mass-matrix}), can be brought to 
quasi-Dirac form $\widehat{m}_{0}$, given by Eq.\,(\ref{mass-matrix1}) with real parameters $m$ and $\epsilon$,
by a certain U(2) transformation $V$,
\begin{equation} 
\widehat m_{0} = V^{T}\widehat{m}V\,.
\label{mass-matrix2}
\end{equation} 
Indeed, 6 real parameters in the matrix $\widehat{m}$ are diminished to 2 in $\widehat m_{0}$ by 
4 parameters of U(2) rotations. 

In the limit $\epsilon=0$ the neutron becomes a Dirac particle, and the baryon symmetry 
${\rm U}(1)_{\cal B}$  associated with SU(2) rotations around 3-rd axis with 
 a diagonal generator $\tau^3\!/2$ arises. Non-zero Majorana mass 
$\epsilon$ breaks this symmetry but 
in real situation $\epsilon \ll m$, the neutron behaves practically as Dirac particle, 
and  $U(1)_{\cal B}$ remains an approximate symmetry.\footnote{Present experimental 
limits on $n-\bar n$ oscillation \cite{Phillips:2014fgb} yield the upper bound $\epsilon < 2.5 \times 10^{-33}$~GeV. }
It is convenient to discuss discrete symmetries in this basis. 
\\[1mm]

\noindent 
{\bf 5.} 
In the Weyl description with the mass matrix $\widehat m_{0}$ given by (\ref{mass-matrix1}) the charge conjugation {\bf C},
\begin{equation}\label{UC}
{\rm \bf C:} \quad \psi^{1\,\alpha}\longleftrightarrow \psi^{2\,\alpha}, \quad
 \xbar{\psi}_{1\dot\alpha} \longleftrightarrow \xbar{\psi}_{2\dot\alpha} \,,
\end{equation} 
is just interchanging fields of the same chirality but with the opposite baryon charges.
In terms of U(2) transformations it is the SU(2) rotation by angle $\pi$ around the first axis
up to the factor $(-i)$ which is the U(1) rotation:
\begin{equation}\label{UC_1}
{\rm \bf C:} \quad \psi \to {U_C}\, \psi, \quad 
U_C={\rm e}^{-i\pi/2}{\rm e}^{i\pi\tau^{1}/2}=\tau^{1}
\, . 
\end{equation} 
This is in the basis where the mass matrix has the quasi-Dirac form (\ref{mass-matrix1}).\footnote{ In the 
limit $\epsilon =0$, when only 3-rd axis is fixed, $m= {\rm Re}\, \mu^3$, 
one can consider any combination of rotations around 1-st and 2-nd axes, 
$U_C=\cos\omega\,  \tau^1 + \sin\omega\, \tau^2$. This is the origin of the well-known 
phase freedom in the definition of  {\bf C} transformation for Dirac fermion,  
$n \to {\rm e}^{-i\omega} n^c$, $n^c \to {\rm e}^{i\omega} n$. Non-zero $\epsilon$ 
removes the phase freedom and leaves only the possibility $\sin\omega=0$. }
For generic form of the mass matrix we can use Eq.\,({\ref{mass-matrix2}) to get
\begin{equation}\label{UC1}
U_C=V\tau^{1}V^{\dagger}\,.
\end{equation} 

Moreover, we can write the matrix  $U_C$ in an arbitrary basis,
\begin{equation}\label{UCa}
\begin{split}
&U_C={\exp}(-i\pi/2)\exp(i\pi\tau^{a}n^{a}/2)= n^{a}\tau^{a}\,, \\[1mm]
&n^{a}=\frac{\epsilon^{abc}\,{\rm Im}\,\mu^{b}\,{\rm Re}\,\mu^{c}}{\big|\epsilon^{abc}\,{\rm Im}\,\mu^{b}\,{\rm Re}\,\mu^{c}\big|}
\, ,
\end{split}
\end{equation} 
with a straightforward geometrical interpretation. Indeed, it is just a combination of the SU(2) rotation around 
 the normal $n^{a}$ to the plane of ${\rm Re}\,\mu^{a}$ and ${\rm Im}\,\mu^{a}$ by angle $\pi$ with the
 chiral U(1) rotation by angle $(-\pi/2)$. Evidently, this is a discrete symmetry of the mass matrix,
 $U_C^T \widehat m U_C = \widehat m$\,. The SU(2) rotation by $\pi$ changes the sign of $\widehat m $
 and the U(1) rotation, $\exp(-i\pi/2)=-i\,$, compensates this sign.

 The transformation {\bf C} together with ${\bf  C}^{2}=I$ composes the discrete $Z_{2}$ subgroup 
 that survives from U(2) for a generic mass term. 
 The only other discrete symmetry is $Z_{2}$ associated with changing sign for 
all fermion fields.

Let us turn now to the parity
transformation {\bf P}$_{z}$  defined by by Eq.\,(\ref{pz}) in terms of Dirac spinors. 
In terms of Weyl spinors (\ref{doubl}) and (\ref{doubl-r}) 
the inversion of space coordinates then implies 
\begin{equation}\label{UaPz} 
\begin{split}
 {\rm \bf P}_z\!: ~~~\psi^{1\alpha} \to i\xbar \psi_{2\dot \alpha}\,, \quad \psi^{2\alpha} \to i\xbar \psi_{1\dot \alpha},\\[1mm]
 \xbar\psi_{1\dot\alpha } \to i\psi^{2\alpha}\,, \quad \xbar\psi_{2\dot\alpha } \to i\psi^{1\alpha}  \, . 
 \end{split}
\end{equation} 
This is in the basis where the mass matrix has the form (\ref{mass-matrix1}). 
Again, similar to {\bf C}, it can be written  
in the form:
\begin{equation}\label{UPz} 
{\bf P}_z\!: ~~ \psi \to i\xbar \psi \, \,U_{\!P}  \,,
\quad  \xbar\psi \to i\,U_{\!P}^\dagger \, \psi\,,
\end{equation} 
where $U_{\!P}=\tau^{1}$  in the basis (\ref{mass-matrix1}) and $U_{\!P}=V\,\tau^{1}\,V^{T}$
in an arbitrary basis.
The transformation (\ref{UPz}) clearly demonstrates \mbox{${\bf P}_{z}^{2}=-1$}.\footnote{In the limit 
$\epsilon =0$, {\bf P}$_z$ transformation can be combined with U(1)$_{\cal B}$ rotations. In particular, 
{\bf P} transformation (\ref{parity}) is a combination of {\bf P}$_z$ and  a discrete baryon rotation 
$i\tau^3 = \exp(i\pi\tau^3/2)$, so that {\bf P}$^2=1$ and ${\rm \bf CP} = - {\rm \bf PC}$. Once again, 
such a definition of parity makes sense only for a Dirac fermion.}

The operation ${\bf CP}_{\!z}$ which changes both, charge and chirality, has the form,
\begin{equation}\label{CPz} 
{\bf  CP}_z\!: ~ \psi \to  i \xbar \psi \,U_{\!C}^{\dagger}U_{\!P}\!=\! i\xbar \psi\, V V^{T}, ~
\xbar\psi \to U_{\!P}^{\dagger}   U_{\!C}\psi \! =\!iV^{*}V^{\dagger}\psi\, .
\end{equation} 
It is just  $\psi^{k\alpha} \to i \xbar{\psi}_{k\dot\alpha}\,$
 and  $\xbar{\psi}_{k\dot\alpha} \to i \psi^{k\alpha} $ in the basis (\ref{mass-matrix1}).
 Note, that {\bf C} and ${\bf P}_{\!z}$ commute and 
 $({\bf  CP}_z)^{2}=-1$.

Finally, one can define {\bf T}  transformation which besides the time inversion and reordering 
operators in the Lagrangian (\ref{weyl})
implies  
\begin{equation}\label{Tz} 
{\bf T}\!: \quad \psi^{i \alpha} \to  \xbar \psi_{i\dot \alpha}, \quad \xbar \psi_{i\dot\alpha} \to
- \psi^{i\alpha}\, , 
\end{equation} 
in the basis (\ref{mass-matrix1}) and
\begin{equation}\label{Tz} 
{\bf T}\!: \quad \psi \to  \xbar \psi\, V V^{T}, \quad \xbar \psi \to
- V^{*}V^{\dagger}\psi^{i\alpha}\, , 
\end{equation} 
in the arbitrary basis. Clearly, {\bf T} anti-commutes with {\bf CP}$_{\!z}$ and 
 ${\rm \bf T}^2=-1$. 
 
Combining, we get $ {\bf CP}_z{\rm \bf T}$
transformation, which acts as  $n\to i\gamma^5 n$ on the Dirac spinor
together with inversion of all space-time coordinates and reordering of 
 operators in the Lagrangian,
 \begin{equation}\label{CPT} 
{\bf CP}_z{\bf T}\!: \quad \psi^{1\alpha} \to i \psi^{i\alpha}, \quad
\xbar\psi_{i\alpha} \to -i \xbar\psi_{i\alpha}
\, .
\end{equation} 
It satisfies $({\rm \bf CP}_z{\rm \bf T})^2=-1$ and presents an invariance of any
local and Lorentz-invariant Lagrangian.

Concluding this section, let us emphasize that we have shown that for any pattern of the neutron mass terms, 
including the Dirac mass respecting the baryon number conservation, as well as the Majorana ones violating 
it by two units, one can always consistently define 
the operations of parity transformation {\bf P}$_z$ and charge conjugation {\bf C} 
as preserved symmetries in spite of breaking of  the baryon charge conservation.
 In fact, a generic mass matrix 
$\widehat m$ in (\ref{mass-matrix}) can be always rotated by flavor transformation 
$V^T \widehat m V$ to a pseudo-Dirac form (\ref{mass-matrix1}) 
where these symmetries are defined in an unique way.

Thus, the neutron-antineutron oscillation in itself does not violate discrete symmetries. 
However, {\bf C}, {\bf P}$_z$ 
and also {\bf CP}$_z$ (which is an equivalent of {\bf T}), generically will not be respected by the interaction terms. 
Consider, e.g., the neutron $\beta$-decay $n \to p e\bar{\nu}$, implying that interaction 
has the standard, baryon charge preserving, form. Then the presence of 
${\cal L}_{\not{\cal B}}$ terms would induce also  the ``wrong" decays 
$n \to \bar{p} \,e^+ \nu$ (though extremely suppressed). Furthermore, {\bf CP}$_z$ 
violation could be manifested in difference of branching ratios of ``wrong" decays 
between the neutron and antineutron, 
${\rm Br}(n\! \to \!\bar{p} \, e^+ \nu) \neq {\rm Br}(\bar n\! \to\! p\, e^+ \bar\nu)$, 
even if observation of these decays is only a {\it gedanken} possibility. 
However, some {\bf CP}$_z$ violating processes related to new ${\cal B}$-violating 
physics that induces the $n-\bar n$ oscillation can be at the origin of the baryon asymmetry 
of the Universe. 
\\[1mm]

\noindent
{\bf 6.} 
In the Standard Model (SM) conservations of baryon ${\cal B}$ and lepton ${\cal L}$ numbers  
are related to accidental global symmetries of the SM Lagrangian.\footnote{Nonperturbative
breaking of ${\cal B}$ and ${\cal L}$, preserving ${\cal B}-{\cal L}$, is extremely small.}
The violation of ${\cal B}$ by two units can originate only from new physics beyond SM which 
would induce the effective six-quark interaction
\begin{equation}
\begin{split}
&{\cal L}\,(\Delta {\cal B}=-2)=\frac{1}{M^{5}}\sum c_{i} {\cal O}^{i} \,,\\
& {\cal O}^{i}= T^{i}_{A_{1}A_{2}A_{3}A_{4}A_{5}A_{6}}q^{A_{1}}q^{A_{2}}q^{A_{3}}q^{A_{4}}q^{A_{5}}q^{A_{6}}\,,
\end{split}
\label{D9}
\end{equation}
where coefficients $T^{i}$ account for different flavor, color and spinor structures and 
 the large mass scale $M$ coming from new physics leads to the  smallness of baryon violation.

In particular, the $n\bar n$ mixing term (\ref{deltaB}) 
emerges as a matrix element between $n$ and $\bar n$ states 
of the operator (\ref{D9}), see diagram in Fig.\,\ref{fig1},
\begin{equation} 
\langle \bar n \vert \,{\cal L}\,(\Delta {\cal B}=-2)\,\vert n\rangle= -\frac 1 2 \,\epsilon \,v_{\bar n}^{T}C\,u_{n}\,,
\label{matel}
\end{equation}
where $u_{n}$, $v_{\bar n}$ are Dirac spinors for $n$, $\bar n$. Generically, it gives a complex value 
for $\epsilon$ but  by a phase redefinition of $n,\,\bar n$ states we always can make it real and positive.
Thus, an estimate of the parameter $\epsilon$, which is inverse of the oscillation time $\tau_{n\bar n}$, is
\begin{equation}
\epsilon =\frac{1}{\tau_{n\bar n}}\sim \frac{\Lambda_{\rm QCD}^6}{M^5}\,.
\end{equation}
\begin{figure}[t]
\begin{center}
\includegraphics[width=6cm]{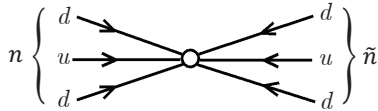}
\caption{
\label{fig1}
Diagram for generating  $n - \bar n$ mixing terms }
\end{center}
\end{figure}
%%%%%%%%%%%%%%%%%%%%%%%%%%
%%%%%%%%%%%%%%%%%%%%%%%%%%%
\begin{figure}[t]
\vspace{-0.7cm}
\begin{center}
\includegraphics[width=6cm]{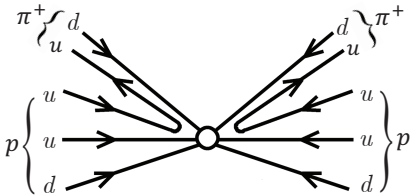}
%\vskip -3.5cm
\caption{
\label{fig2}
Inducing  $pp \to \pi^+ \pi^+$ annihilation via operators (\ref{D9}) }
\end{center}
\end{figure}
%%%%%%%%%%%%%%%%%%%%%%%%%%

For $u$ and $d$ quarks of the first generation the full list of $\Delta {\cal B}=-2$ six-quark operators was determined 
in Refs.\,\cite{Rao,Caswell},
\begin{equation}
\begin{split}
&{\cal O}^{1}_{\chi_{1}\chi_{2}\chi_{3}} =u^{iT}_{\chi_{1}}Cu^{j}_{\chi_{1}} d^{\,kT}_{\chi_{2}}Cd^{\,l}_{\chi_{2}} d^{\,mT}_{\chi_{3}}Cd^{\,n}_{\chi_{3}}\big[\epsilon_{ikm}\epsilon_{jln}+\\
&\qquad\qquad\quad\epsilon_{ikn}\epsilon_{jlm}+\epsilon_{jkm}\epsilon_{nil}+\epsilon_{jkn}\epsilon_{ilm}\big],\\[1mm]
&{\cal O}^{2}_{\chi_{1}\chi_{2}\chi_{3}}=u^{iT}_{\chi_{1}}Cd^{j}_{\chi_{1}} u^{\,kT}_{\chi_{2}}Cd^{\,l}_{\chi_{2}} d^{\,mT}_{\chi_{3}}Cd^{\,n}_{\chi_{3}}\big[\epsilon_{ikm}\epsilon_{jln}+\\
&\qquad\qquad\quad\epsilon_{ikn}\epsilon_{jlm}+\epsilon_{jkm}\epsilon_{nil}+\epsilon_{jkn}\epsilon_{ilm}\big],\\[1mm]
&{\cal O}^{3}_{\chi_{1}\chi_{2}\chi_{3}}=u^{iT}_{\chi_{1}}Cd^{j}_{\chi_{1}} u^{\,kT}_{\chi_{2}}Cd^{\,l}_{\chi_{2}} d^{\,mT}_{\chi_{3}}Cd^{\,n}_{\chi_{3}}\big[\epsilon_{ijm}\epsilon_{kln}+\\
&\qquad\qquad\quad\epsilon_{ijn}\epsilon_{klm}\big].
\end{split}
\end{equation}
Here $\chi_{i}$ stands for $L$ or $R$ quark chirality. Accounting for relations
\begin{equation}
\begin{split}
&{\cal O}^{1}_{\chi L R}\!={\cal O}^{1}_{\chi R L}\,,\quad {\cal O}^{2,3}_{L R\chi}\!={\cal O}^{2,3}_{R L\chi}\,,\\[1mm]
&{\cal O}^{2}_{\chi \chi \chi^{\prime}}-{\cal O}^{1}_{\chi \chi \chi^{\prime}}\!=3{\cal O}^{3}_{\chi \chi \chi^{\prime}}\,,
\end{split}
\end{equation}
we deal with 14 operators for $\Delta {\cal B}=-2$ transitions and 14 Hermitian conjugated ones 
for $\Delta {\cal B}=+2$.

The ${\rm \bf P}_{z}$ reflection interchanges 
$L$ and $R$ chirality $\chi_{i}$ in the operators $O^{i}_{\chi_{1}\chi_{2}\chi_{3}}$. Note, that
the ${\rm \bf P}_{z}$ reflection for $u$ and $d$ quarks is defined similar to the neutron by Eq.\,(\ref{pz}). 
This is consistent with the $udd$ wave function of neutron. Thus, we can divide operators into 
${\rm \bf P}_{z}$ even and ${\rm \bf P}_{z}$ odd ones,
\begin{equation}
O^{i}_{\chi_{1}\chi_{2}\chi_{3}}\pm L \leftrightarrow R\,.
\end{equation}

The charge conjugation {\bf C} transforms operators $O^{i}_{\chi_{1}\chi_{2}\chi_{3}}$ into the Hermitian conjugated $[O^{i}_{\chi_{1}\chi_{2}\chi_{3}}]^{\dagger}$. Again, our phase definitions for quarks are consistent with those for neutron.
So, combinations
\begin{equation} 
O^{i}_{\chi_{1}\chi_{2}\chi_{3}\!}\pm{\rm H.c.}
\end{equation}
represent {\bf C}  even and {\bf C} odd operators. 
In total, we break all 28 operators into four groups with different ${\rm \bf P}_{z}$, {\bf C} and ${\rm \bf CP}_{z}$
features, each group contains seven operators,
\begin{equation}
\begin{split}
&\big[O^{i}_{\chi_{1}\chi_{2}\chi_{3}}\!\!+\! L \leftrightarrow R\big] \!+\!{\rm H.c.},~~ {\rm\bf P}_{z}\!=+\,,~{\rm\bf C}=+\,,~{\rm\bf CP}_{z}=+\,;\\[1mm]
&\big[O^{i}_{\chi_{1}\chi_{2}\chi_{3}}\!\!+\! L \leftrightarrow R\big] \!-\!{\rm H.c.},~~ {\rm\bf P}_{z}\!=+\,,~{\rm\bf C}=-\,,~{\rm\bf CP}_{z}=-\,;\\[1mm]
&\big[O^{i}_{\chi_{1}\chi_{2}\chi_{3}}\!\!-\! L \leftrightarrow R\big] \!+\!{\rm H.c.},~~ {\rm\bf P}_{z}\!=\,-,~{\rm\bf C}=+\,,~{\rm\bf CP}_{z}=-\,;\\[1mm]
&\big[O^{i}_{\chi_{1}\chi_{2}\chi_{3}}\!\!-\! L \leftrightarrow R\big] \!-\!{\rm H.c.},~~ {\rm\bf P}_{z}\!=\,-,~{\rm\bf C}=-\,,~{\rm\bf CP}_{z}=+\,.
\end{split}
\end{equation}
Only the first seven operators,
which are both ${\rm\bf P}_{z}$ and {\bf C} even,  contribute to $n\bar n$ oscillations.
It is,  of course, up to small corrections due to electroweak 
interactions where the discrete symmetries are broken. 

What about the remaining 21 combinations which are odd either under ${\rm \bf P}_{z}$  or {\bf C} transformations?
Although they do not contribute to the $n-\bar n$ transition, their effect show up in instability of nuclei.
This source of instability in this case is not due to neutron-antineutron oscillations but due to 
processes of annihilation of two nucleons inside nucleus
like $N+N \to \pi+\pi$, and, in particular, two proton annihilation,
$p\,p \to \pi^+\pi^+$, shown on Fig.\,\ref{fig2}. This could be particularly interesting in case 
of suppressed $n\bar n$ oscillations.
%\\[1mm]

%\noindent
%{\bf 8.}
The operators of the type of (\ref{D9}) involving strange quark, like $udsuds$, could induce $\Lambda-\bar\Lambda$  mixing. However, such operators  would also lead  to nuclear instability via nucleon annihilation into kaons $N + N \to K + K$, see the diagram in Fig.\,\ref{fig2} where 
in upper lines $d$ quark is substituted by $s$ quark (and $\pi^{+}$ by $K^{+}$). 
In fact, nuclear instability bounds on 
$\Lambda-\bar\Lambda$ mixing are only mildly, within an order of magnitude, weaker 
than with respect to $n-\bar n$ mixing which makes hopeless the possibility to detect 
 $\Lambda-\bar\Lambda$ oscillation in the hyperon beam. 
(Instead, it can be of interest to search for the nuclear 
decays into kaons  in the large volume detectors.)
The nuclear instability limits on $\Lambda-\bar\Lambda$ mixing are 
about 15 orders of magnitude stronger than the sensitivity 
$\delta_{\Lambda\bar\Lambda}\sim 10^{-6}$~eV which can be achieved in the 
laboratory conditions \cite{Kang}. The nuclear stability limits make hopeless also 
the laboratory search of $bus$-like baryon oscillation due to operator 
$usbusb$ suggested in Ref.\,\cite{Kuzmin}.  
\\[1mm]

\noindent
{\bf 7.} 
Our above consideration refers to the neutron-anti\-neutron oscillation in vacuum.
Now we show that even in the presence of magnetic field no new $|\Delta {\cal B}|=2$ operator
appears. A similar consideration was done in Ref.\,\cite{Voloshin:1987qy}  in application to 
a possible magnetic moment of neutrino.

In the Weyl formalism the field strengths tensor $F_{\mu\nu}$ is substituted by the symmetric 
tensor $F_{\alpha\beta}$ and its complex conjugate $\xbar F_{\dot\alpha\dot\beta}$\,. They correspond
to $\vec E\pm i \vec B$ combinations of electric and magnetic fields. Then Lorentz invariance allows only
two structures involving electromagnetic fields,
\begin{equation}
F_{\alpha\beta}\psi^{i\alpha}\psi^{k\beta}\epsilon_{ik}\,,\quad \xbar F_{\dot\alpha\dot\beta}\bar\psi_{i}^{\dot\alpha}\bar\psi_{k}^{\dot\beta}\epsilon^{ik}\
\end{equation}
Antisymmetry in flavor indices implies that spinors with the opposite baryon charges enter. 
So both operators preserve the baryon charge, 
and in fact they describe interactions with the magnetic and electric dipole moments of the neutron. 
In terms of Majorana mass eigenstates (\ref{Maj}), these are transitional moments between 
$n_1$ and $n_2$. However, no transitional moment can exist between $n$ and $n^c$. 

The authors of Ref.\,\cite{Gardner:2014cma} realize that the operator $n^{T}\sigma^{\mu\nu} CnF_{\mu\nu}$ with $\Delta {\cal B}= -2$ is vanishing 
due to Fermi statistics. They believe, however, that a composite nature of neutron changes the situation and a new type of magnetic moment in $\Delta {\cal B}=\pm 2$ transitions may present. In other words, they think that the effective Lagrangian description is broken for composite particles. 

To show that is not the case let us consider the process of annihilation of two neutrons into virtual photon,
\begin{equation}
n(p_{1})+n(p_{2}) \to \gamma^{*}(k)\,,
\label{nngamma}
\end{equation}
which is the crossing channel to $n-\bar n \gamma^{*}$ transition.  The number of invariant amplitudes for
the process (\ref{nngamma}) which is $1/2^{+}+1/2^{+}\to 1^{-}$ transition is equal to one.
Only orbital momentum $L=1$ and total spin $S=1$ in the two neutron system are allowed by angular 
momentum conservation and Fermi statistics. The gauge-invariant form of the  amplitude is
\begin{equation}
u^{T}\!(p_{1})C\gamma^{\mu}\gamma_{5}u(p_{2}) \,k^{\nu}\big(k_{\mu}\epsilon_{\nu}-k_{\nu}\epsilon_{\mu}\big),
\label{nngamma1}
\end{equation}
where $u_{1,2}$ are Dirac spinors describing neutrons and $\epsilon_{\mu}$ refers to the gauge potential.
In space representation we deal with $\partial^{\nu}F_{\mu\nu}$ the quantity which vanishes outside of the source of 
the electromagnetic field, and, in particular, for the distributed magnetic field. It proves that there is no place for magnetic moment of $n-\bar n$ transition, and effective 
Lagrangian description does work. Let us also remark that $n \to \bar n \gamma^{*}$ transition 
with a virtual photon connected to the proton, as well as $nn \to\gamma^{*}$ 
annihilation, would destabilise the nuclei even in the absence of $n-\bar n$ mass mixing. 

Even in the absence of new $n-\bar n$ magnetic moment the authors of \cite{Gardner:2014cma} 
claim that suppression of $n-\bar n$ oscillations by external magnetic field can be overcome
by applying the magnetic field transversal to quantization axis. 
Following our criticism \cite{BV}  Gardner and Yan recognized  in \cite{Gardner}
that it would break the rotational invariance. As a consequence the magnetic field suppression does present 
indeed. 

The situation is different if  one considers oscillation $n-n'$ where $n'$ is a mirror 
neutron,  twin of the neutron from hidden mirror sector  \cite{Berezhiani:2005hv}. 
In this case one deals with the mass mixing  between two Dirac fermions, $\varepsilon \ov{n} n' + {\rm h.c.}$, conserving a combination of baryon numbers ${\cal B} +  {\cal B}'$.  
Hence, also operators  $\overline{n}\sigma^{\mu\nu} n' F_{\mu\nu}$ and  
$\overline{n}\sigma^{\mu\nu} \gamma^5 n' F_{\mu\nu}$ are allowed which describe respectively the transitional 
magnetic and electric dipole moments between $n$ and $n'$ states (and the similar operators with mirror 
electromagnetic field,  $F_{\mu\nu} \to F'_{\mu\nu}$).  Since underlying new physics generating 
$n-n'$ mixings generically should violate CP-invariance, both transitional magnetic and electric dipole 
moments can be of the same order. 
In large enough magnetic (or electric) field $n \to n'$ 
 transition probabilities should not depend on the field value,  
 with possible implications for the search of neutron$-$mirror neutron transitions, 
 and in particular for testing experimentally solution of the neutron lifetime puzzle  
 via $n-n'$ transitions \cite{puzzle}. 
\\[1mm]

\noindent
{\bf 8.} Our use of the effective Lagrangian for the proof means that the Lorentz invariance and {\bf CPT}
are crucial inputs. Once constraints of Lorentz invariance are lifted new $|\Delta {\cal B}|=2$ operators could show up. 

Such operators were analyzed in Ref.\,\cite{Babu:2015axa} for putting limits on the Lorentz invariance breaking.  
In particular, the authors suggested the operator $n^{T}C\gamma^{5}\gamma^{2}n$ as an example
which involves spin flip and, correspondingly, less dependent on magnetic field surrounding.

Note, however, that besides breaking of Lorentz invariance this operator breaks also 3d rotational invariance, i.e.,
isotropy of space. Such anisotropy could be studied by measuring spin effects in neutron-antineutron transitions.\\[1mm]

\noindent
{\bf 9.}
The construction we used for neutron-antineutron transition could be applied to mixing of
massive neutrinos. As an example, let us take the system of left-handed $\nu_{e}$ and
$\nu_{\mu}$ and their conjugated partners, right-handed $\bar\nu_{e}$ and
$\bar\nu_{\mu}$. One can ascribe them \cite{Konopinski:1953gq} a flavor charge ${\cal F}={\cal L}_{e}-{\cal L}_{\mu}$ (analog of ${\cal B}$), to be (+1) for $\nu_{e}$  and (-1)
for $\nu_{\mu}$. Then, {\bf C} conjugation is interchange of $\nu_{e}$ and
$\nu_{\mu}$. Again, ${\cal F}$ breaking mass term would be {\bf C} and ${\rm \bf P}_{z}$ even but odd for {\bf P}.

A similar scenario can be staged in case of Dirac massive neutrino. \\[1mm]

\noindent
{\bf 10.} In summary, we show that the Lorentz and {\bf CPT} invariance lead to the unique 
$|\Delta {\cal B}|=2$ operator in the effective Lagrangian for
the neutron-antineutron mixing. This mixing is even under the charge conjugation {\bf C} as well as under the modified parity 
${\rm \bf P}_{\!z}$ which takes the same value $i$ for both, neutron and antineutron in contrast with standard $(+1)$ and $(-1)$ values. It means that observation of the neutron-antineutron mixing {\em per se} does not give a signal of {\bf CP} violation.
It could be compared with the $K^{0}-\xbar K^{0}$ transition amplitude with $|\Delta S|=2$ where to separate  {\bf CP}  conserving and {\bf CP} breaking parts one needs to relate it to  $|\Delta S|=1$ decay amplitudes.

We applied the discrete symmetries to classification of possible $|\Delta {\cal B}|=2$ six-quark operators,
separating those which contributes to the neutron-antineutron mixing. Other $|\Delta {\cal B}|=2$ operators 
contribute to instablity of nuclei.

We also showed that switching on external magnetic field  influences the level splitting,
what suppresses $n-\bar n$ oscillations, but does not add any new $|\Delta {\cal B}|=2$ operator in contradistinction 
with recent claims in literature.

Our classification of $|\Delta {\cal B}|=2$ operators coming
from new physics, could be useful in association with Sakharov conditions for baryogenesis which involves both,
non-conservation of baryon charge and CP-violation.
\\[2mm]

We thank Martin Einhorn, Susan Gardner, Yuri Kamyshkov, Kirill Melnikov, Rabi Mohapatra, Adam Ritz and Misha Voloshin for helpful discussions.
A.V. appreciates hospitality of the Kavli Institute for Theoretical Physics where his research was supported 
in part by the National Science Foundation under Grant No.\ NSF PHY11-25915.

%\end{document}  

 \end{document}